\documentclass{PoS}
\usepackage{amsfonts,amssymb,amsmath,amsthm,cite}
\newtheorem{assumption}{Assumption}[section]
\newtheorem{theorem}[assumption]{Theorem}
\newtheorem{corollary}[assumption]{Corollary}

\newtheorem{lemma}[assumption]{Lemma}
\newtheorem{definition}[assumption]{Definition}
\newtheorem{prop}[assumption]{Proposition}

\newcommand{\B}{\mathcal{B}}              
\newcommand{\C}{\mathbb{C}}              
\newcommand{\R}{\mathbb{R}}              
\newcommand{\CC}{\mathcal{C}}              

\newcommand{\CB}{\mathcal{B}}

\newcommand{\acts}{\triangleright}

\newcommand{\cop}{\triangle}                 
\newcommand{\ts}{\otimes}                 

\newcommand{\bq}{\begin{eqnarray}}
\newcommand{\eq}{\end{eqnarray}}
\newcommand{\be}{\begin{equation}}
\newcommand{\ee}{\end{equation}}
\newcommand{\enq}{\nonumber\end{eqnarray}}

\title{$\kappa$-Minkowski differential calculi and star product}
\ShortTitle{$\kappa$-Minkowski differential calculi and star product}
\author{Flavio Mercati \\
        Departamento de F{\'i}sica Te{\'o}rica, Universidad de Zaragoza, Zaragoza 50009, Spain \\
        E-mail: \email{flavio.mercati@gmail.com}}
\author{\speaker{Andrzej Sitarz}\thanks{Partially supported by the Polish Government grant 1261/7.PRUE/2009/7} \\
Institute of Physics, Jagiellonian University, Reymonta 4, 30-059 Krak\'ow, Poland \\
E-mail:\email{sitarz@if.uj.edu.pl}}

\abstract{We present the star-product algebra of the $\kappa$-deformation 
of Minkowski space and the formulation of Poincar\'e covariant differential
calculus. We use these tools to construct a twisted $K$-cycle over the algebra 
and a twisted cyclic cocycle.}

\FullConference{Corfu Summer Institute on Elementary Particles and Physics - Workshop on Noncommutative Field Theory and Gravity, September 8-12, 2010, Corfu Greece}

\begin{document}

\section{Introduction}\label{intr}
The $\kappa$-deformation of Minkowski space originated from the deformation
of Poincar\'e symmetry, which was introduced by Lukierski, Nowicki, Ruegg 
\cite{LNR}. The first realisation of the algebra was, however, constructed 
on the usual commutative algebra of functions. The correct and precise 
formulation of the $\kappa$-Minkowski space and its symmetries appeared later, 
introduced by Majid and Ruegg \cite{MajRue} together with the discussion 
of the bicrossproduct structure of the $\kappa$-Poincar\'e.

The $\kappa$-Minkowski space is a quantum homogeneous space of $\kappa$-deformed 
Poincar\'e symmetries. Treated alone as an algebra it could be described as 
a nilpotent Lie algebra with $d$ self-adjoint generators $x_i$, $i=0, \ldots d-1,$ 
and relations:
\begin{align}
\label{kappa_Mink}
[x_0,\,x_i]= \tfrac{i}{\kappa} \,x_i, \quad [x_i,\,x_j]=0, \quad i,j=1,\dots,d-1,
\end{align} 
where $\kappa > 0$ can be viewed as a deformation parameter. Formally, the 
commutative limit of the coordinate algebra of Minkowski space is obtained 
in the limit $\kappa \to \infty$. 

The differential calculi on $\kappa$-Minkowski were thoroughly studied by several authors but almost exclusively in the purely algebraic setup (see \cite{MKJ} and
the references therein). In this paper we want to present the construction of 
the fully covariant differential calculus \cite{AS} on the star-product version of the $\kappa$-Minkowski algebra recently introduced by Durhuus and Sitarz \cite{DS}.

For simplicity we shall restrict ourselves to $d=2$ case. 

\section{Star product}

The early attempts to describe the algebra of the $\kappa$-Minkowski space through 
a star-product date to
Agostini \cite{Ago}, followed recently by the series of papers of Dabrowski and 
Piacitelli \cite{DabPia1,DabPia2,DabPia3,DabGodPia}. A family of star products 
in form of a formal series expansions were considered in \cite{MS1}.
In a most recent work, Durhuus and Sitarz \cite{DS} obtained a much simpler 
realisation of the star product, together with precise description of the algebra, 
its properties and the symmetries.

We shall briefly sketch here here the main results of that paper, which provides
a convenient realisation of the $\kappa$-Minkowski algebra. This opens several 
possibilities towards further study of geometric properties of $\kappa$-Minkowski
within the noncommutative geometry.

Let $\CB$ be the linear space of functions over $\R^2$, which are Fourier transforms 
of functions with a compact support. We define the $\kappa$-star product as:

\be
\label{star2}
(f*g)(\alpha,\beta) =  \frac{1}{2\pi}  \int  dv \int du \, 
f(\alpha+\frac{u}{\kappa},\beta) g(\alpha, e^{-v} \beta) e^{-i u v}
\ee

or, equivalently as:

\be
\label{star3}
(f*g)(\alpha,\beta) =  \frac{1}{2\pi}  \int  dv \int du \, 
f(\alpha+u,\beta) g(\alpha, e^{-\frac{v}{\kappa}} \beta) e^{-i u v}.
\ee

The algebra $\B$ comes equipped with the involution:
\be\label{inv2}
f^*(\alpha,\beta) = \frac{1}{2\pi} \int dv\int du\, \bar{f}(\alpha+u, e^{-\frac{v}{\kappa}} \beta) 
e^{-iuv}. 
\ee

For detailed proofs of associativity and other properties, including the derivation of formulas 
(\ref{star2}-\ref{inv2}) we refer the reader to \cite{DS}. 

The above definition is the Moyal-type formulation of a product over the subspace of 
smooth functions on $\R^2$. To see that it corresponds exactly to $\kappa$-Minkowski
one needs to extend the product to a much bigger algebra, which includes polynomials,
as it has been demonstrated in \cite{DS}.

If $\CC$ denotes smooth functions of two variables, which are polynomially bounded, and,
additionally they are Fourier transforms in the first variable of tempered distributions
with compact support (in this variable) then the star product is still well defined. To see
it let us denote by $\tilde{f}$ a Fourier transform of a function over $\R^2$ in the first
variable. Then, from (\ref{star3}) we can rewrite the star product as:

\be
\label{star4}
f*g(\alpha,\beta) =  \frac{1}{\sqrt{2\pi}}  \int  dv \, 
\tilde{f}(v,\beta) g(\alpha, e^{-\frac{v}{\kappa}} \beta) e^{i \alpha v}.
\ee

The integral makes sense, since $\tilde{f}$ is of compact support in $v$, in fact,
it makes sense even if $\tilde{f}$ is a distribution, provided that its support 
remains a compact subset of $\R$.

Since both $i_\alpha(\alpha,\beta) = \alpha$ and $i_\beta(\alpha,\beta)=\beta$ are
of polynomial bound and are Fourier transforms of delta Dirac distribution in the first
variable (which is clearly of compact support), then it makes sense to calculate their
products. Using (\ref{star4}) and the standard calculus of distributions we obtain:

$$
(i_\alpha * i_\beta)(\alpha,\beta) = \frac{i}{\kappa} \beta + \alpha \beta, \;\;\;\; 
(i_\beta * i_\alpha)(\alpha,\beta) = \beta \alpha,
$$
and therefore
$$
[i_\alpha,i_\beta] =  \frac{i}{\kappa} i_\beta,
$$
so we can identify $i_\alpha$ with $t$ and $i_\beta$ with $x$.

The main advantage of the star product formulation is that, unlike in almost
all considerations so far, we do not need to work exclusively with polynomials 
or apply rather inelegant procedure of normal ordering of coordinates. Moreover, 
we can make perfect sense of expressions, which are not of polynomial type. 

\section{The $\kappa$-deformed Poincar\'e algebra}

The $\kappa$-deformation was treated from its very origins as a homogeneous space of
deformed $\kappa$-Poincar\'e algebra \cite{MajRue,Zak}. 

For that reason in addition to the star-product formulation we need to consider the
realisation of the deformed symmetries, which form a Hopf algebra. We follow the 
notation of \cite{MajRue}.

We recall here the generators of 2-dimensional $\kappa$-deformed Poincar\'e algebra. 

$$
\begin{aligned}
&[P,E] = 0, \;\;\;\;\;  &[N,E] = P, \\
&\cop P = P \ts 1 + e^{i \frac{E}{\kappa}} \ts P,  \;\;\;\;\;  &\cop E = E \ts 1 + 1 \ts E,
\end{aligned}
$$
and for the relations involving the boost,
$$
\begin{aligned}
&[N,P]=  \frac{i\kappa}{2}(1-e^{ 2i \frac{E}{\kappa}}) + \frac{i}{2\kappa}P^2,\;\;\;\;\;
& \cop N = N \ts 1 + e^{i \frac{E}{\kappa}} \ts N.
\end{aligned}
\label{boost}
$$

To make sense of the algebra as a Hopf algebra (as we want to avoid using the exponential 
function of one of the generators) we formally introduce an operator ${\cal E}$ (and its 
inverse) and rewrite the relations accordingly.

$$
\begin{aligned}
&[P,{\cal E}] = 0, \;\;\;\;\; & [E, {\cal E}] = 0, \;\;\;\;\; & \\
&[N, {\cal E}] = \frac{i}{\kappa} {\cal E} P, \;\;\;\;\; 
&[N, P] = \frac{i \kappa}{2}(1-{\cal E}^2) + \frac{i}{2\kappa}P^2, & \\
&\cop P = P \ts 1 + {\cal E} \ts P,  \;\;\;\;\;   
&\cop N = N \ts 1 + {\cal E} \ts N,  \;\;\;\;\;
&\cop {\cal E} = {\cal E} \ts {\cal E}. 
\end{aligned}
$$

The action of these generators on the generators of the $\kappa$-Minkowski algebra is
the same as in the classical case:

$$ 
\begin{aligned} 
&P \acts t = 0, \;\;\;\;\;\; & P \acts x  = 1, \\
&E \acts t = 1, & E \acts x  = 0, \\
&N \acts t = -x, & N \acts x  = -t, \\
&{\cal E} \acts t = t + \frac{i}{\kappa}, & {\cal E} \acts x  = x.
\label{actionrel}
\end{aligned}
$$

Our notation here differs slightly from \cite{DS} as we wanted to keep the relations
(\ref{actionrel}) without any change, simply replace $E$ and $P$ above by $-iE$ and $-iP$
to obtain the relations from \cite{DS}.

To proceed further we need to know how to extend the action of the generators of 
Poincar\'e algebra to the star-product algebra $\CB$.

\begin{definition}
Let $f \in \B$ so that it is a Fourier transform of $\hat{f} \in C_0(\R^2)$. 
We define a one parameter group of linear operations on $\B$ in the following 
way. For any $\gamma \in \R$ let:
\be
T_\gamma (f)(\alpha,\beta) = \frac{1}{2\pi} \int du dv \, \hat{f}(u,v) e^{\gamma u} 
e^{-i(\alpha u +\beta v)}. 
\label{tgamma}
\ee
\end{definition}

Since $\hat{f}$ is a function on $\R^2$ with compact support, so 
is its product by $e^{\gamma u}$. Hence the definition is well posed 
and the obtained function is in $\B$. It is easy to verify that
$$ T_\gamma T_\delta = T_{\gamma+\delta}. $$

The explicit formula for the map $T_\gamma: f \to T_\gamma f$ is:
\be 
T_\gamma(f)(\alpha,\beta) =  \frac{1}{2\pi} \int dp ds \, f(s,\beta) e^{-\gamma p} e^{-ip(s-\alpha)}. \label{transl}
\ee

Observe that we can formally write
\be
(T_\gamma f)(\alpha,\beta) = f(\alpha+i\gamma, \beta). 
\label{itrans}
\ee

If $\tilde{f}$ denotes the Fourier transform of $f$ with respect to the first variable, it is
also useful to have the following formula:

\be
\widetilde{(T_\gamma f)}(v,\beta) = \tilde{f}(v,\beta) e^{\gamma v}. 
\label{tg}
\ee

Of course, for (\ref{transl}) to make sense as an integral we need to be careful of the
order of integration, similarly one can make exact sense of (\ref{itrans}) - for precise
mathematical statements see \cite{DS}.

One can claim (see Proposition 4.1 in \cite{DS})
\begin{prop}
The map $T_\gamma: f \to T_g f$ is an algebra automorphism.
\end{prop}

\begin{proof}
We present here a sketch of the proof (again, for details see \cite{DS}). 
Using the formula for the star product and the explicit formula for the transformation
$T_\gamma$ we have:

$$
\begin{aligned}
T_\gamma(f)&*T_\gamma(g)(\alpha,\beta) = 
\frac{1}{2\pi} \int du dv \, T_\gamma(f)(\alpha+u,\beta) T_\gamma(g)(\alpha,e^{-v} \beta) e^{-iuv} \\
&= \frac{1}{(2\pi)^3} \int du dv \int dp ds \int dp' ds' \,
f(s,\beta) g(s',e^{-v} \beta)  e^{-\gamma p} e^{-\gamma p'} e^{-ip(s-\alpha-u)} e^{-ip'(s'-\alpha)} e^{-iuv} \\
&= \frac{1}{(2\pi)^2} \int dv ds \int dp' ds' \,
f(s,\beta) g(s',e^{-v} \beta)  e^{-\gamma (p'+v)} e^{-iv(s-\alpha)} e^{-ip'(s'-\alpha)} = \ldots 
\end{aligned}
$$
changing now variables: $s=S+U$, $s'=S$, $P=p'+v$ we obtain:
$$
\begin{aligned}
\ldots &= \frac{1}{(2\pi)^2} 
\int dU dv \int dP dS\, f(S+U,\beta) g(S,e^{-v} \beta)  
e^{-\gamma P} e^{-iP(S-\alpha)} e^{-iUv} \\
&= \frac{1}{(2\pi)} \int dP dS\, (f*g)(S,\beta) e^{-\gamma P} e^{-iP(S-\alpha)} 
= T_\gamma \left(f*g \right)(\alpha,\beta).
\end{aligned}
$$
\end{proof}

We have:

\begin{theorem}[see Theorem 4.2 in \cite{DS}]
The algebra $\B$ is a Hopf module algebra with respect to the 
following action of the momentum algebra, with the generators 
$E,P,{\cal E}$ represented as linear operators on $\B$.

$$ (E \acts f)(\alpha,\beta) = \frac{\partial}{\partial \alpha}, \;\;\;
   (P \acts f)(\alpha,\beta) = \frac{\partial}{\partial \beta}, \;\;\;
   ({\cal E} \acts f)(\alpha,\beta) = (T_{\frac{1}{\kappa}}f)(\alpha,\beta).
$$
\end{theorem}

\begin{proof}
The only thing to verify is the compatibility of the action of the generators with the
product in the algebra $\B$. Since ${\cal E}$ is group-like and $T_\gamma$ is an
algebra automorphism this part is already done. As the coproduct for $E$ is classical
(Lie-algebra type) and $E$ acts by partial derivation, it is easy to see that this
is also satisfied. It remains only to check the action of $P$:

We calculate:
$$
\begin{aligned}
(P \acts (f&*g))(\alpha,\beta) =  \frac{\partial}{\partial \beta} 
\left( \frac{1}{2\pi} \int du dv f(\alpha+u,\beta) g(\alpha, e^{-\frac{v}{\kappa}} \beta) e^{-iuv} \right) \\
&=  \frac{1}{2\pi} \int du dv \left( \left(\frac{\partial}{\partial \beta} f(\alpha+u,\beta) \right)  g(\alpha, e^{-\frac{v}{\kappa}} \beta)
+ f(\alpha+u,\beta) \left( \frac{\partial}{\partial \beta} g(\alpha, e^{-\frac{v}{\kappa}} \beta) \right) \right) e^{-iuv} \\
&=   ((P \acts f)*g)(\alpha,\beta) + 
\frac{1}{2\pi} \int du dv \, f(\alpha+u,\beta) e^{-\frac{v}{\kappa}} \left( \frac{\partial g}{\partial \beta} \right) (\alpha, e^{-\frac{v}{\kappa}} \beta)  e^{-iuv} \\
&=   ((P \acts f)*g)(\alpha,\beta) + ({\cal E} \acts f)*(P \acts g)(\alpha,\beta),
\end{aligned}
$$
where in the last line we have used:
$$ 
\begin{aligned}
(({\cal E} \acts f) &* g)(\alpha,\beta) = \frac{1}{\sqrt{2\pi}} \int dv \,
 \widetilde{({\cal E} \acts f)}(v,\beta) g(\alpha, e^{-\frac{v}{\kappa}}\beta) e^{i\alpha v} \\
&=  \frac{1}{\sqrt{2\pi}} \int dv \, \tilde{f}(v,\beta) e^{-\frac{v}{\kappa}} g(\alpha, e^{-\frac{v}{\kappa}} \beta) e^{i\alpha v} \\
&=  \frac{1}{2\pi} \int du dv \, f(\alpha+u,\beta) e^{-\frac{v}{\kappa}} g(\alpha, e^{-\frac{v}{\kappa}} \beta) e^{-iuv}.
\end{aligned}
$$
\end{proof}

\section{Poincar\'e invariant differential calculus}
In what follows we define the first and higher order differential calculus on the
$\kappa$-deformed star-product algebra defined in (\ref{star2}).

\begin{lemma}
We can rewrite the action of the external derivative in the bicovariant 
and Lorentz invariant differential calculus, described in \cite{AS} 
in the following way:
\be
d f = dx \, ({\cal E}^{-1} P \acts f) 
       - i \frac{\kappa}{2} \psi_+ \, \left(( {\cal E}^{-1} -1) \acts f \right)
       +  \frac{1}{2} \psi_- \, \left(( \frac{i}{\kappa}{\cal E}^{-1}P^2 
       + i \kappa ({\cal E} - 1)) \acts f \right),
\label{defd}
\ee
where the bimodule rules between the functions and the generating one-forms are:
\be
\begin{aligned}
& f \, dx = dx \, f + \psi_+ \, \frac{i}{\kappa}(P \acts f), \\
& f\, \psi_+ = \psi_+ \,({\cal E}^{-1} \acts f) 
+ dx \, \frac{2i}{\kappa} ({\cal E}^{-1} P \acts f) 
- \psi_- \, \frac{1}{\kappa^2} ({\cal E}^{-1} P^2 \acts f), \\
& f \, \psi_- = \psi_- \, ({\cal E} \acts f).
\end{aligned}
\label{birel}
\ee
\end{lemma}
The proof comes from direct computations of the invariance with respect to 
the action of $N$. To see that one recovers the calculus of \cite{AS}, one
obtains explicitly for the generators we have:

\be
\begin{aligned}
x \, dx = dx \, x + \frac{i}{\kappa} \psi_+ \\
t \, dx = dx \,t, \\
x \,\psi_+ = \psi_+ \,x + 2 \frac{i}{\kappa} dx, \\
t \,\psi_+ = \psi_+ \,t + \frac{i}{\kappa} \psi_+, \\
x \,\psi_- = \psi_- \,x, \\
t \,\psi_- = \psi_- \,t - \frac{i}{\kappa} \psi_-,
\end{aligned}
\ee

and we can easily see that one recovers the three-dimensional bicovariant
and Poincar\'e covariant calculus.

\section{Differential algebra and $K$-cycle}

Following the standard procedure how to extend the first order bicovariant 
differential calculus to higher order forms we obtain (compare with \cite{AS}) 
that the generating one-forms  $dx, \psi_+, \psi_-$ are all exact and anticommute 
with each other.

Our aim here, will be to extend the differential algebra to the $K$-cycle (see 
\cite{Connesbook} for details) - that is to equip $\Omega^*(\CB)$ with a closed 
graded trace. 

From \cite{DS} we know that there exists a unique twisted trace on the 
algebra, which is invariant under the action of the Poincar\'e algebra:

\begin{prop}[see Proposition 4.7 in \cite{DS}]
Let $\int f$ denote the standard Lebesgue integral over $\R^2$ for any 
$f \in \CB$. Then it has the following properties:
\begin{itemize}
\item $\int$ is a twisted trace, that is for all $f,g \in \CB$:
$$ \int \, (f*g) = \int \left( T_{\frac{1}{\kappa}}(g)*f \right), $$
\item  $\int$ is invariant with respect to the action of $\kappa$-Poincar\'e, 
that is, for any $h \in {\cal P}_\kappa$ 
and $f \in \CB$:
$$ \int \, (h \acts f) = \varepsilon(h) \int \, f. $$
\end{itemize}
\end{prop} 
\begin{proof} 
The proof is by a straightforward computation, for further details we refer to 
\cite{DS}. Let us take $f,g \in \CB$. We have:

$$
\begin{aligned}
\int f*g =& \int d\alpha d\beta \, (f*g)(\alpha,\beta) =  
\frac{1}{2\pi}  \int d\alpha d\beta  \int  dv \int du \, 
f(\alpha+u,\beta) g(\alpha, e^{-\frac{v}{\kappa}} \beta) e^{-i u v} \ldots
\end{aligned}
$$

On the other hand, using (\ref{transl}):
$$
\begin{aligned}
\int (T_{\frac{1}{\kappa}}(f))*g =& \int d\alpha d\beta \, 
((T_{\frac{1}{\kappa}}(f))*g)(\alpha,\beta) =   \\
=& \frac{1}{(2\pi)^2} \int d\alpha d\beta  \int  dv \int du 
\int dp ds \, g(s,\beta) e^{-\frac{1}{\kappa} p} e^{-ip(s-\alpha-u)} 
f(\alpha, e^{-\frac{v}{\kappa}} \beta) e^{-i u v} \\
=& \frac{1}{2\pi} \int d\alpha d\beta \int dv ds 
g(s,\beta) e^{-\frac{1}{\kappa} v} f(\alpha, e^{-\frac{v}{\kappa}} \beta) 
e^{-iv(s-\alpha)} = \ldots
\end{aligned}
$$
where by a simple change of variables: $s = a$, $\alpha = a+r$, $v=-q$:
$\beta = -\frac{v}{\kappa} b$ we obtain:  
$$
\ldots = \frac{1}{2\pi} \int da db \int dq dr 
f(a+r, b) g(a, e^{-\frac{1}{\kappa}} q) e^{-iqr} 
$$
which is the same expression as above, which ends the proof.
\end{proof}

Using this twisted trace we can construct the twisted K-cycle over the algebra $\B$:

\begin{prop}
Let $\omega$ be a three-form in the Poincar\'e covariant differential calculus over $\CB$:
$$ \omega = (dx \wedge \psi_+ \wedge \psi_-) \, f, $$
where $f \in \CB$. The following linear functional from $\Omega^3(\CB) \to \C$:
\be
\int \, \omega = \int \, f.
\ee
is a closed graded twisted trace over the differential algebra $\Omega^*(\CB)$.
\end{prop}

\begin{proof}
First, let us show that it is closed. Let $\omega = d \rho$, where $\rho$ is a
two-form:
\be
 \rho = (dx \wedge \psi_+) \, f_+ + (dx \wedge \psi_-) \, f_- +
(\psi_+ \wedge \psi_-) \, f_{+-}. 
\label{2form}
\ee

To see that the integral of $d\rho$ vanishes it is sufficient to observe that 
the integral of each of the component functions of $d\rho$ vanishes. This, however
is the consequence of the Poincar\'e invariance of the trace $\int$. Since the
counit of the algebra elements, which define the action of the external derivative 
$d$ are:
\be
\varepsilon\left( {\cal E}^{-1} P \right) = 0, \;\;\;
\varepsilon\left( {\cal E}^{-1} -1 \right) = 0, \;\;\;
\varepsilon\left( \frac{i}{\kappa}{\cal E}^{-1}P^2 + i \kappa ({\cal E} - 1) \right) = 0. 
\ee
Then the trace of their actions on any element of the algebra must vanish.
To see the property of the twisted trace, first for the product between a three-form $\omega = (dx \wedge \psi_+ \wedge \psi_-) \, g$ and a function $f$, we have:

\be
\begin{aligned}
\int \omega * f &= \int (dx \wedge \psi_+ \wedge \psi_- \, g*f)  \\
&=  \int (g*f) = \int ( T_\frac{1}{\kappa}(f)*g ) =  \int dx \wedge \psi_+ \wedge \psi_- \, 
( T_\frac{1}{\kappa}(f)*g ) \\
& = \int T_\frac{1}{\kappa}(f)* (dx \wedge \psi_+ \wedge \psi_- \, g ) = \int T_1(f)*\omega.
\end{aligned}
\ee

For the product of a one form with a two form observe that the action of $T_\gamma$
on the generating one-forms is trivial:

$$ T_\gamma(dx) = dx, \;\;\;\; T_\gamma(\psi_+) = \psi_+, \;\;\;\; T_\gamma(\psi_-) = \psi_-. $$
 
Indeed, using the formula (\ref{defd}) and the fact that $T_\gamma$ is group like 
and commutes with $P$ and ${\cal E}$ we get the above property immediately. 

Take now any two-form $\rho$, as in (\ref{2form}). Then, using the relations
(\ref{birel}) we have:

$$
\begin{aligned}
\rho \wedge dx & = (dx \wedge \psi_+ \wedge \psi_-) f_{+-}, \\
\rho \wedge \psi+ & = - (dx \wedge \psi_+ \wedge \psi_-) \left( {\cal E}^{-1} \acts f_{+} \right), \\
\rho \wedge \psi- & =   (dx \wedge \psi_+ \wedge \psi_-) \left( {\cal E} \acts f_{-} \right).
\end{aligned}
$$

Since the twisted trace is invariant under the action of ${\cal E}$ we have:

$$
\begin{aligned}
\int \rho \wedge dx & = \int dx \wedge \rho = \int T_{\frac{1}{\kappa}}(dx) \wedge \rho, \\
\int \rho \wedge \psi+ & = \int \psi_+ \wedge \rho = \int T_{\frac{1}{\kappa}}(\psi_-) \wedge \rho, \\
\int \rho \wedge \psi- & = \int \psi_- \wedge \rho = \int T_{\frac{1}{\kappa}}(\psi_-) \wedge \rho.
\end{aligned}
$$
which, together with the twisted trace property for the wedge product of 
a three form and a function completes the proof.
\end{proof}

As a consequence of the above construction we can present an explicit 
analytic formula for the three-cyclic cocycle over the $\kappa$-Minkowski 
algebra.

\begin{corollary}
Let $f_0,f_1,f_2,f_3 \in \CB$ be the functions over $\kappa$-Minkowski space.
The multilinear functional:
$$ \phi(f_0,f_1,f_2,f_3) = \int f_0 \, df_1 \wedge df_2 \wedge df_3, $$
defines a twisted cyclic cocycle over $\CB$.
\end{corollary}

This property is an immediate consequence of the theorem (2.1) of \cite{KMT}. 
The explicit formula for $\phi$ (which could be derived easily using the definition
and bimodule relations for the differential calculus and the closed graded trace) 
is in no way enlightening, so we skip it. 

\section{$\kappa$-Minkowski and Rieffel's deformation.}

Although $\kappa$-deformation of the Minkowski space has been obtained by
a procedure which is quite different from the Moyal deformation, it is
interesting to notice that thanks to the star-product formulation we can
find many similarities. In this section we shall show how to write it using
integral presentation of deformation by the action of $\R^2$ in Rieffel's \cite{Rie}
approach.

First, let us start with the following action of $\R^2$ on the algebra $\CB$:
$$ \eta_{(r,s)}(f) (\alpha,\beta) = f(\alpha + r, e^{-s} \beta), \;\;\; (r,s) \in \R^2. $$

Further, define a linear operator $J: \R^2 \to \R^2$:
$$ J (r,s) = (s, 0). $$
Unlike in the Rieffel's case, the operator is not invertible, in fact, it is nilpotent.
However, nothing forbids us from defining the following $J$-star product on $\CB$:
$$
(f*_{J} g)(\alpha,\beta) = \frac{1}{(2\pi)^2} \int du_1 du_2 dv_1 dv_2 \, 
(\eta_{J(u_1,u_2)}(f))(\alpha,\beta) (\eta_{(v_1,v_2)}(g))(\alpha,\beta) 
e^{-i(u_1 v_1+u_2 v_2)}.  
$$

The involution on the space $\CB$ can also be written in the language of Rieffel deformation:

$$ f^*(\alpha,\beta) = \int du_1 du_2 \, 
   \eta_{u_1,u_2}(\bar{f})(\alpha,\beta) e^{-i (u_1, u_2) J (u_1, u_2)}.$$

We have:

\begin{lemma}
If $f,g \in \CB$ then the $*_{J}$ product is well-defined and equals the star product (\ref{star2}).
\end{lemma}

\begin{proof}
Indeed, calculating:
$$
\begin{aligned}
(f *_J g)(\alpha,\beta) = & \frac{1}{(2\pi)^2} 
\int du_1 du_2 dv_2 dv_2 \, (\rho_{J(u_1,u_2)}(f))(\alpha,\beta) 
(\rho_{(v_1,v_2)}(g))(\alpha,\beta) e^{-i(u_1 v_1+u_2 v_2)}. \\
  =& \frac{1}{(2\pi)^2} \int du_1 du_2 dv_2 dv_2 \, 
     f(\alpha+u_2,\beta) g(\alpha+v_1,e^{-v_2} \beta) e^{-i(u_1 v_1+u_2 v_2)} \\
  =& \frac{1}{(2\pi)} \int du_2 dv_2 \, f(\alpha+u_2,\beta) g(\alpha,e^{-v_2} \beta) e^{-i u_2 v_2}
\end{aligned}
$$
\end{proof}

As an immediate corollary we obtain that the $\R^2$ action on $\CB$ 
is actually an algebra automorphism:

\be
\begin{aligned}
\rho_{(r,s)}(f*g)(\alpha,\beta) &= 
\frac{1}{2\pi} \int f(\alpha+r+u,e^{-s} \beta) g(\alpha+r,e^{-v-s}b) e^{iuv} du dv \\ & = 
\left( \eta_{(r,s)}(f) \right) * \left( \eta_{(r,s)}(g) \right)(\alpha,\beta).
\end{aligned}
\ee

\section{Conclusions}

We have demonstrated here that the geometry of the $\kappa$-deformation can be 
approached from a different angle, when presented through the star product 
formulation. The most important result is the construction of the $K$-cycle,
which leads to the first known example of a twisted cyclic cocycle over 
$\kappa$-Minkowski.

We see the construction we provided as the first step towards analysis of 
Connes-type geometry in the $\kappa$ case. Of course, this raises a series
of interesting questions. First of all, we do not know whether the twisted 
cyclic cocycle is not trivial. If this is the case, then, surprisingly, 
$\kappa$-deformation will be (in view of cyclic cohomology) of dimension $3$. 
Furthermore, the geometric interpretation of the constructed cocycle will 
be a challenge: in particular, to find a Dirac operator and a spectral 
triple (or twisted spectral triple), which provide the geometric realisation of 
the cocycle.

So far, the analysis of $\kappa$-deformations was based on the polynomial algebra
or formal expressions, like in \cite{AACADA,GAC}. Using the star product in this
formulation and its symmetries we open new possibilities to go beyond that approach, 
which might use Connes' noncommutative geometry to study geometric properties 
of that noncommutative space. 


\end{document}